\documentclass[aps,twocolumn,amsmath,amssymb,floatfix]{revtex4}
\usepackage{graphicx}
\usepackage{dcolumn}
\usepackage{bm}
\usepackage{amsfonts}
\usepackage{graphicx}% Include figure files
\usepackage{dcolumn}% Align table columns on decimal point
\usepackage{bm}% bold math
\usepackage{amsmath}% needed for subequations
\usepackage{amssymb}

\begin{document}
\title{Optical conductivity of multi-Weyl semimetals}
\author{Seongjin Ahn$^{1}$}
\author{E. J. Mele$^2$}
\email{mele@physics.upenn.edu}
\author{Hongki Min$^{1,2}$}
\email{hmin@snu.ac.kr}
\affiliation{$^1$ Department of Physics and Astronomy, Seoul National University, Seoul 08826, Korea}
\affiliation{$^2$
Department of Physics and Astronomy, University of Pennsylvania, Philadelphia, Pennsylvania 19104, USA}

\date{\today}

\begin{abstract}
Multi-Weyl semimetals are new types of Weyl semimetals which have anisotropic non-linear energy dispersion and a topological charge larger than one, thus exhibiting a unique quantum response. Using a unified lattice model, we calculate the optical conductivity numerically in the multi-Weyl semimetal phase and in its neighboring gapped states, and obtain the characteristic frequency dependence of each phase analytically using a low-energy continuum model. The frequency dependence of longitudinal and transverse optical conductivities obeys scaling relations that are derived from the winding number of the parent multi-Weyl semimetal phase and can be used to distinguish these  electronic states of matter. 
\end{abstract}

\maketitle
%\normalsize

%%%%%%%%%%%%%%%%%%%%%%%%%%%%%%%%%%%%%%%%%%%%%%%%%%%%%%%%%%%%%%%%%%%%%%%%%%%%%%%%%%%%%%%%%%%%%%%%%%%%
{\em Introduction}.
A Weyl semimetal (WSM) is a gapless topological state of matter possessing $\bm k$-space singularities where its valence and conduction bands make contact at a point. This singularity is a $\bm k$-space monopole providing a quantized source or sink of a Berry's flux and can occur only in materials in which either time reversal symmetry or inversion symmetry is broken. In the prototypical WSM, a twofold band degeneracy at the Weyl point is broken linearly in momentum in all directions and the node is characterized by the topological winding number (also referred to as chirality) $\pm 1$. A transition to an insulating phase is possible only if Weyl nodes with opposite chirality pairwise merge and annihilate. The gapped phase produced  by this merger can be in a normal insulating state or a topological quantum anomalous Hall state.  The linear dispersion around the Weyl point has important consequences for the low frequency optical properties, which have been explored theoretically and used as an experimental fingerprint of the topological state \cite{Wan2011,Hosur2012,Ashby2014,Sushkov2015,Chen2015a,Tabert2016a,Tabert2016b,Xu2016,Neubauer2016}.

A $\bm k$-space merger of Weyl points with the {\it same} chirality produces a new type of Weyl semimetal, referred to as a multi-Weyl semimetal (m-WSM) \cite{Xu2011,Fang2012}. In these states, the merger of the nodes is robust if it is protected by a point group symmetry. The low energy dispersion can then be characterized by double (triple) Weyl nodes with linear dispersion along one symmetry direction and quadratic (cubic) dispersion along the remaining two directions. Because of the change in topological nature, the enhancement of the density of states, the anisotropic non-linear energy dispersion and a modified spin-momentum locking structure, these states will have optical and transport signatures  that distinguish them from elementary WSMs.

In this Rapid Communication, we report calculations of the optical conductivity in m-WSMs, and analyze their characteristic frequency dependence in the semimetallic state and in nearby insulating states, focusing on the effects of the winding number, lattice regularization and phase transitions. 
%We study how the optical conductivity changes for these different phases using a unified lattice model in which the model parameters are varied, and obtain the characteristic frequency dependence analytically using a low energy continuum model.  
We find that the results for m-WSMs can be clearly distinguished from 
those for 
%the low-frequency electrodynamics of elementary 
WSM’s by their low-energy frequency dependence, which is determined by the winding number of the m-WSM phase.

%%%%%%%%%%%%%%%%%%%%%%%%%%%%%%%%%%%%%%%%%%%%%%%%%%
\begin{figure}[htb]
\includegraphics[width=0.95\linewidth]{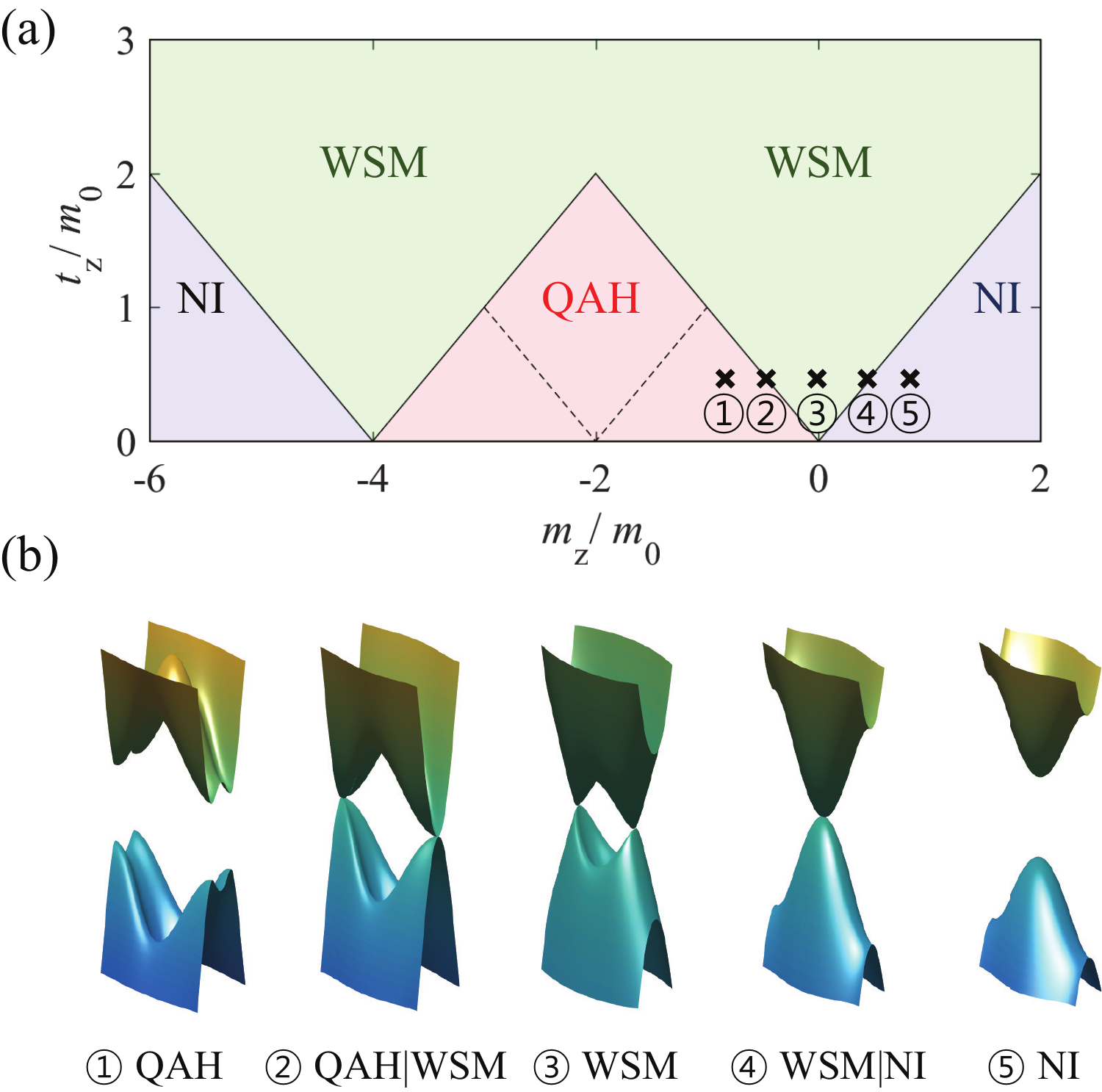}
\caption{
(a) Phase diagrams of $J=2$ lattice models on the $t_z/m_0$ and $m_z/m_0$ plane and (b) evolution of the energy band structure from the 3D quantum anomalous Hall (QAH) phase to the normal insulator (NI) phase. Here, we use several values of $m_z/m_0$ corresponding to different phases, indicated by circled numbers in the phase diagram. QAH$|$WSM and WSM$|$NI denote the transition phase between 3D QAH and WSM, and WSM and NI, respectively. The phase diagram for $J=1$ has a similar shape, but has a different phase boundary between the WSM and 3D QAH represented by the dashed line \cite{Chen2015b}.}
\label{fig:phase_diagram_J2}
\end{figure}
%%%%%%%%%%%%%%%%%%%%%%%%%%%%%%%%%%%%%%%%%%%%%%%%%%

%%%%%%%%%%%%%%%%%%%%%%%%%%%%%%%%%%%%%%%%%%%%%%%%%%%%%%%%%%%%%%%%%%%%%%%%%%%%%%%%%%%%%%%%%%%%%%%%%%%%
{\em Model}.
The low-energy effective Hamiltonian for m-WSMs of order $J$ near a single Weyl point can be described by the Hamiltonian:
\begin{equation}\label{eq:wsm_J}
H_J=\varepsilon_0 \left( \tilde{k}_{-}^J \sigma_{+} + \tilde{k}_{+}^J\sigma_{-}\right)+\hbar v_z k_z \sigma_z,
\end{equation}
where $\tilde{k}_{\pm}=k_{\pm}/k_0$ with $k_\pm=k_x\pm i k_y$, $\sigma_\pm={1\over 2}\left(\sigma_x\pm i\sigma_y\right)$, and $\bm{\sigma}$ are Pauli matrices acting in the space of two bands that make contact at the Weyl point. Here, $v_z$ is the effective velocity along the $k_z$ direction, and $k_0$ and $\varepsilon_0$ are material dependent parameters in units of momentum and energy, respectively. For simplicity, we assumed the axial symmetry around the $k_z$-axis. Note that the eigenenergies of the Hamiltonian are given by $\varepsilon_{\pm}=\pm \sqrt{\varepsilon_0^2 \tilde{k}_{\parallel}^{2J}+(\hbar v_z k_z)^2}$ where $\tilde{k}_{\parallel}=\sqrt{\tilde{k}_x^2+\tilde{k}_y^2}$, and the in-plane energy dispersion is characterized by $J$. Thus the winding number determines not only the topological nature of the wave function but also the anisotropic energy dispersion of the system.
	
Let us consider a lattice model that shows at some parameter range the m-WSM phase described by Eq.~(\ref{eq:wsm_J}).
A simple lattice model for the Weyl semimetals with $J=1$ which has inversion symmetry with broken time-reversal symmetry is given by \cite{Delplace2012,Turner2013,Chen2015b}
\begin{eqnarray}
\label{eq:wsm_lattice_J1}
H_1&=& t_x \sin (k_xa) \sigma_x + t_y \sin (k_ya) \sigma_y + M_z\sigma_z, \\
M_z&=&m_z-t_z\cos (k_za)+m_0[2-\cos (k_xa) -\cos (k_ya)] \nonumber,
\end{eqnarray}
where $a$ is the lattice spacing, and $t_{x,y,z}$, $m_z$ and $m_0$ are material dependent parameters. Similarly, we can generalize the above lattice model in Eq.~(\ref{eq:wsm_lattice_J1}) to $J=2$ so that near the Weyl points the low-energy Hamiltonian reduces to the form of Eq.~(\ref{eq:wsm_J}) \cite{Jian2015}:
\begin{eqnarray}
\label{eq:wsm_lattice_J2}
H_2&=& t_x [\cos (k_ya) -\cos (k_xa)] \sigma_x \nonumber \\
&+& t_y \sin (k_xa) \sin (k_ya) \sigma_y + M_z \sigma_z.
\end{eqnarray}
Depending on the model parameters, the Hamiltonian in Eqs.~(\ref{eq:wsm_lattice_J1}) and (\ref{eq:wsm_lattice_J2}) shows various phases such as normal insulators (NIs), Weyl semimetals, and three-dimensional (3D) quantum anomalous Hall (QAH) states, as shown in Fig.~\ref{fig:phase_diagram_J2} along with the corresponding energy band structures. The phase diagram for $J=2$ has a similar shape to that for $J=1$ \cite{Chen2015b}, but because of the change in the electronic structure, the optical properties in the m-WSMs show a strong dependence on their chirality.

For the continuum model corresponding to each phase, we choose the parameter range where Weyl nodes arise at $(k_x,k_y)=(0,0)$. Other choices of parameter ranges give fundamentally identical settings. Using the $\bm{k}\cdot\bm{p}$ method, we can write a generic continuum model for various phases as \cite{Supplemental}
\begin{equation}
\label{eq:generic_continuum_model_mass_term}
H=\varepsilon_0 \left( \tilde{k}_{-}^J \sigma_{+} + \tilde{k}_{+}^J\sigma_{-}\right)+M_z \sigma_z,
\end{equation}
where the mass term is given by $M_z\approx \hbar v_z q_z + \alpha + \beta q_z^2+\gamma (k_x^2+k_y^2)$.
%\begin{equation}
%\label{eq:generic_continuum_model_mass_term}
%M_z\approx \hbar v_z q_z + \alpha + \beta q_z^2+\gamma (k_x^2+k_y^2).
%\end{equation}
%(See App.~\ref{sec:continuum_model} for detailed discussion.)
Here we set $\gamma={m_0 a^2 \over 2}>0$ except for the WSM phase with $\gamma=0$ where the linear term in $M_z$ dominates over the quadratic term associated with $\gamma$ at low energies. 
%Note that for the NI phase and at the transition between the NI and WSM phases (NI$|$WSM), $\alpha=m_z-t_z$ and $\beta={t_z a^2\over 2}$, whereas for the 3D QAH phase and at the transition between the WSM and 3D QAH phases (WSM$|$QAH), $\alpha=m_z+t_z$ and $\beta=-{t_z a^2\over 2}$.
%Thus, for each phase, we find
Note that for the NI (3D QAH) phase, $\alpha=m_z\mp t_z$ and $\beta=\pm{t_z a^2\over 2}$. Then, for each phase, we find
\begin{equation}
\begin{array}{lcccc}
\!\text{NI}\!\! &: & q_z=k_z; & v_z=0; & \alpha,\beta>0, \\
\!\text{NI$|$WSM}\!\! &: & q_z=k_z; & v_z=0; & \alpha=0, \beta>0, \\
\!\text{WSM}\!\! &: & q_z=k_z\mp b; & v_z\neq 0; & \alpha=\beta=0, \\
\!\text{WSM$|$QAH}\!\! &: & q_z=k_z\mp {\pi\over a}; & v_z=0; & \alpha=0, \beta<0, \\
\!\text{QAH}\!\! &: & q_z=k_z\mp {\pi\over a}; & v_z=0; & \alpha,\beta<0,
\end{array}
\end{equation}
where $\cos (ba)\equiv m_z/t_z$ with $|m_z|/t_z <1$. For calculation, we set $k_0=1/a$, $t_x=t_y=4m_0$ and $t_z=0.5 m_0$ with $m_0 >0$, and vary $-m_0<m_z<m_0$ with other parameters fixed to induce various phases.

%%%%%%%%%%%%%%%%%%%%%%%%%%%%%%%%%%%%%%%%%%%%%%%%%%
\begin{figure}[!htb]
\includegraphics[width=1\linewidth]{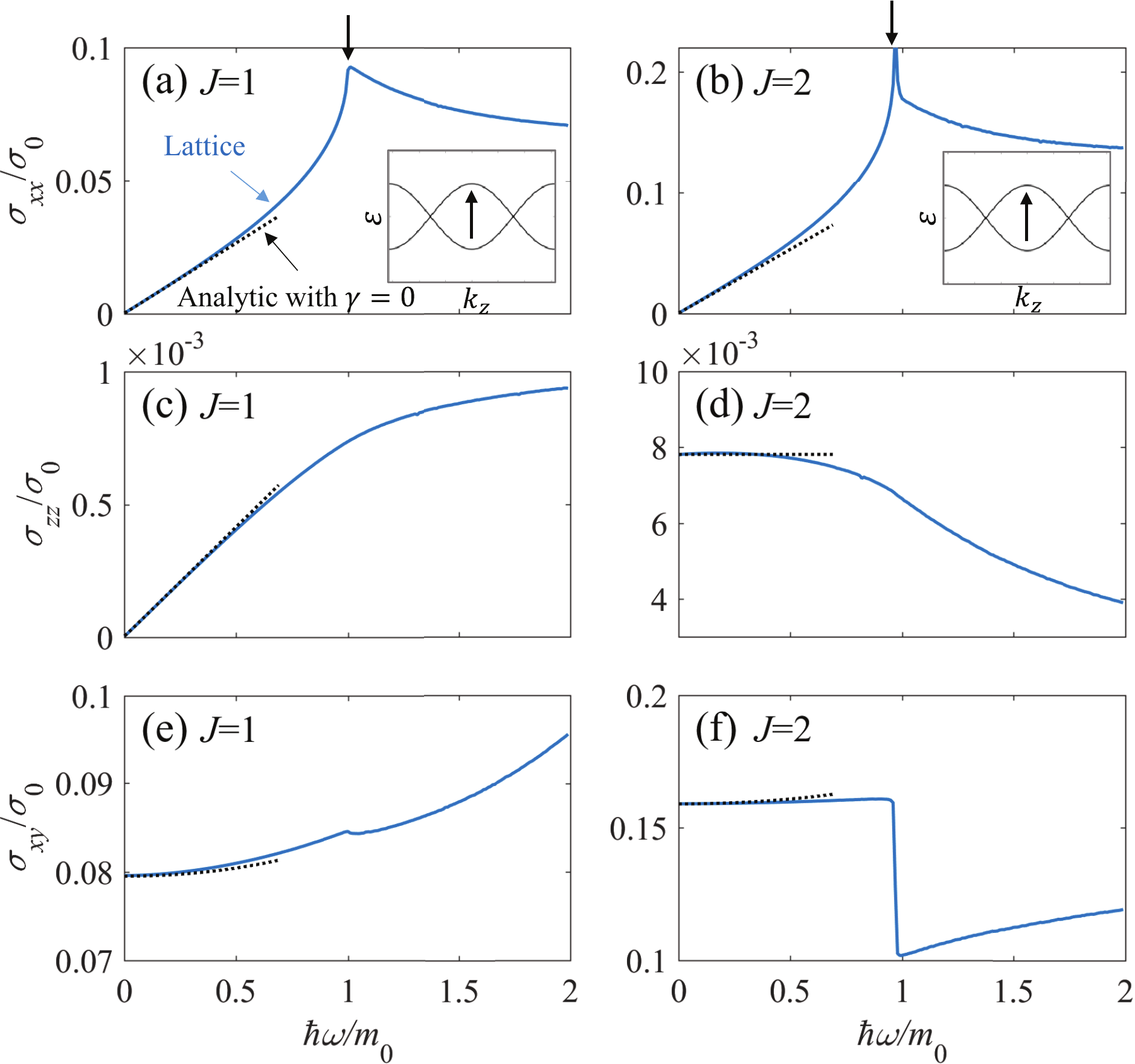}
\caption{
Real part of (a)-(d) longitudinal and (e), (f) transverse optical conductivities in units of $\sigma_0={e^2\over \hbar a}$ for the lattice (blue solid line) and continuum (black dotted line) models in the WSM phase. The arrows in the insets indicate interband transitions corresponding to  kink structures in $\sigma_{xx}(\omega)$ and $\sigma_{xy}(\omega)$. Here, $m_z/m_0=0$, $b=0.5\pi/a$, and $k_{\rm c}=\pi/a$ are used for calculation.
}
\label{fig:WSM_optical_conductivity}
\end{figure}
%%%%%%%%%%%%%%%%%%%%%%%%%%%%%%%%%%%%%%%%%%%%%%%%%%

%%%%%%%%%%%%%%%%%%%%%%%%%%%%%%%%%%%%%%%%%%%%%%%%%%%%%%%%%%%%%%%%%%%%%%%%%%%%%%%%%%%%%%%%%%%%%%%%%%%%
{\em Optical conductivity}.
The Kubo formula for the optical conductivity in the non-interacting limit can be expressed as \cite{Mahan}
\begin{eqnarray}
\label{eq:conductivity}
\sigma_{ij}(\omega)
&=&- \frac{ie^2}{\hbar} \sum_{s,s'} \int \frac{d^3 k}{(2\pi)^3} \frac{f_{s, \bm{k}}-f_{s',\bm{k}}}{\varepsilon_{s,\bm{k}}-\varepsilon_{s',\bm{k}}} \nonumber \\
&\times&\frac{M^{ss'}_i(\bm k)M^{s's}_j(\bm k)}{\hbar\omega+\varepsilon_{s,\bm{k}}-\varepsilon_{s',\bm{k}}+i0^+},
\end{eqnarray}
where $i,j=x,y,z$, $f_{s,\bm{k}}=1/[1+e^{(\varepsilon_{s,\bm{k}}-\mu)/k_{\rm B}T}]$ is the Fermi distribution function, $\mu$ is the chemical potential, and $M^{ss'}_i(\bm k)=\langle{s,\bm{k}}|\hbar\hat{v}_i |{s',\bm{k}}\rangle$, with the velocity operator $\hat{v}_i$ obtained from the relation $\hat{v}_i=\frac{1}{\hbar}\frac{\partial \hat{H}}{\partial  k_i}$.

In the following, we consider only the undoped case with $\mu=0$. 
%The effect of finite $\mu$ simply gives additional features of Pauli blocking for interband transitions and the Drude peak for intraband transitions.
In the clean limit at zero temperature, the real part of the longitudinal optical conductivity for m-WSMs within the continuum model is given by \cite{Supplemental}
\begin{subequations}
\label{eq:WSM_optical_conductivity_longitudinal}
\begin{eqnarray}
\sigma_{xx}(\omega)&=&{g_{\rm N} \over 24\pi}{J e^2\over \hbar v_z}\omega, \\
\sigma_{zz}(\omega)&=&{g_{\rm N} \over 24\pi}{e^2v_z\over\hbar v_\parallel^2}A_{zz}^{\rm WSM}\left(\omega\over\omega_0\right)^{{2\over J}-1}\omega_0,
\end{eqnarray}
\end{subequations}
where $g_{\rm N}=2$ is the number of nodes, $\varepsilon_0=\hbar \omega_0=\hbar v_\parallel k_0$,
$A_{zz}^{\rm WSM}=\frac{3\sqrt{\pi} \Gamma\left({1\over J}\right)}{2^{{2\over J}}J^2\Gamma\left({1\over J}+{3\over 2}\right)}$, and $\Gamma(x)$ is the gamma function \cite{Arfken}.
%For $J=1$, $\left.A_{zz}^{\rm WSM}\right|_{J=1}=1$ and the result in Eq.~(\ref{eq:WSM_optical_conductivity_longitudinal}) reduces to that of conventional Weyl semimetals. 
Note that $\sigma_{xx}(\omega)\propto J\omega$ while $\sigma_{zz}\propto \omega^{{2\over J}-1}$, exhibiting the chirality dependent power-law exponents in frequencies. 
%Also note that the effect of a finite $\mu$ simply produces a small gap due to Pauli blocking and a conventional Drude peak from intraband transitions, which does not alter the characteristic frequency dependence of the conductivity.
Also note that the effect of a finite $\mu$ simply produces a small gap due to Pauli blocking in interband transitions and a conventional Drude peak from intraband transitions, which does not alter the characteristic frequency dependence of the conductivity as long as $\mu$ is not high enough that the effective Hamiltonian is still characterized by a m-WSM Hamiltonian \cite{Supplemental}.

Next, consider the real part of the Hall or transverse optical conductivity.   Note that a sign change of $M_z$ in the Brillouin zone can produce a nontrivial state that supports a Hall effect in the $k_x$-$k_y$ plane for a fixed $k_z$. 
We therefore focus only on the in-plane off-diagonal part $\sigma_{xy}(\omega)$.
If two Weyl nodes with opposite chirality are located at $\pm\bm{b}=\pm b\hat{\bm{z}}$, the real part of the Hall conductivity up to second order in $\omega$ is given by
\begin{equation}
\label{eq:optical_conductivity_transverse}
\sigma_{xy}(\omega)=J\chi\frac{e^2}{\hbar}\left({ b\over 2\pi^2}+\frac{1}{24\pi^2 v_z^2}\frac{b}{k_{\rm c}^2-b^2}\omega^2\right),
\end{equation}
where $k_{\rm c}$ is the cutoff along the $k_z$ direction. Here, $\chi$ represents the right-handed/left-handed chirality, which has $\chi = \pm1$ if the node with positive chirality is at $\pm b\hat{\bm{z}}$ and the other at $\mp b\hat{\bm{z}}$.
Note that the Hall conductivity for m-WSMs is given by $J$ times that for $J=1$ Weyl semimetals, thus their surface states could be manifested by $J$ Fermi arcs connecting the two Weyl nodes.

Figure \ref{fig:WSM_optical_conductivity} shows the calculated optical conductivities for $J=1$ and $J=2$ lattice and continuum models, respectively. At low frequencies, the lattice models are approximated by the corresponding low energy model in Eq.~(\ref{eq:wsm_J}), thus the optical conductivities obtained from the lattice and continuum models are in good agreement. As the frequency increases, however, optical conductivities deviate from the continuum model and show a kink structure in $\sigma_{xx}(\omega)$ and $\sigma_{xy}(\omega)$ at $\hbar\omega=2|m_z-t_z|$ due to the interband transitions between states around the van Hove singularity \cite{Tabert2016b}, as shown in the insets to (a) and (b).

%%%%%%%%%%%%%%%%%%%%%%%%%%%%%%%%%%%%%%%%%%%%%%%%%%
\begin{figure}[htb]
\includegraphics[width=1\linewidth]{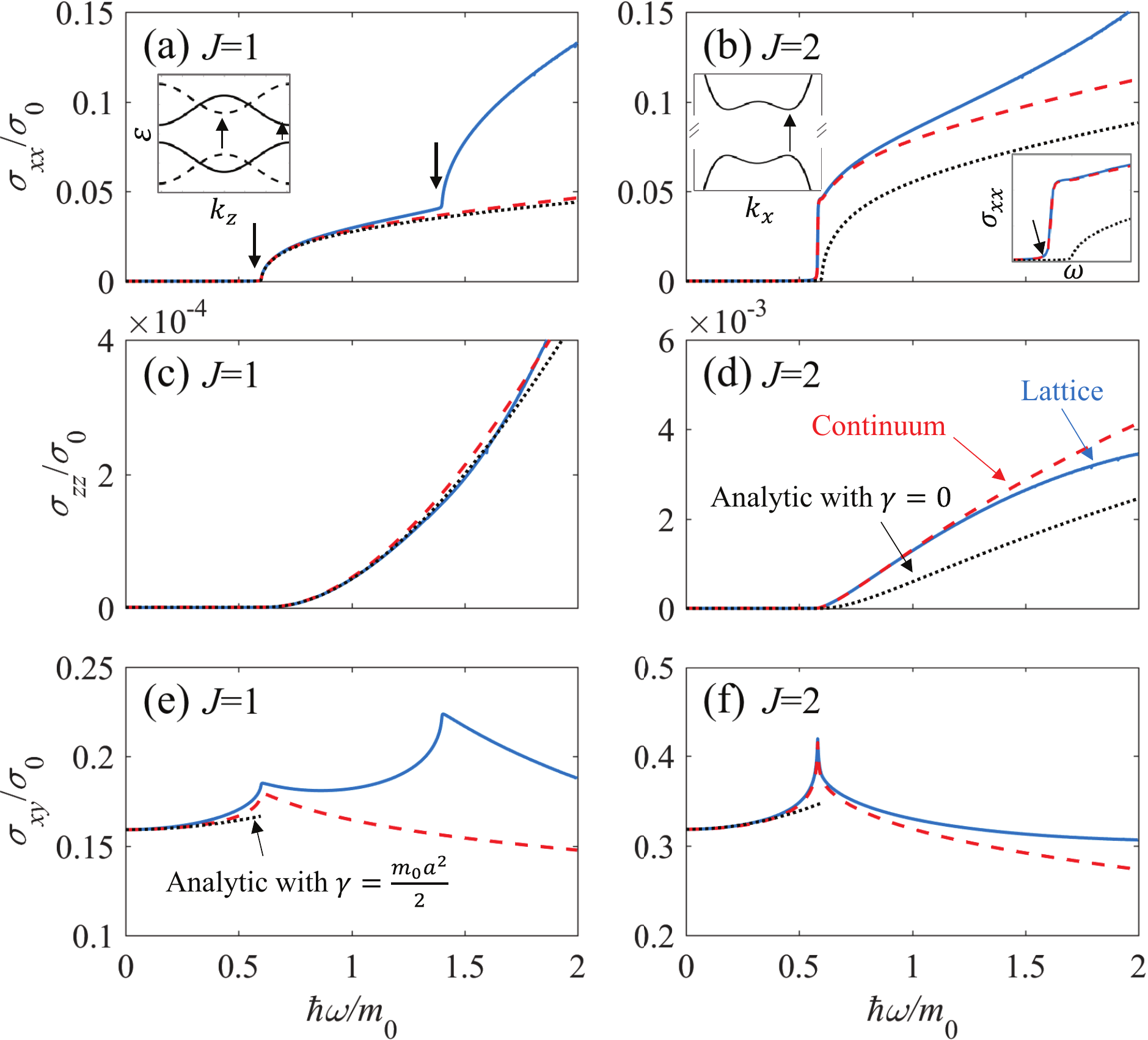}
\caption{
Real part of (a)-(d) longitudinal and (e), (f) transverse optical conductivities in the 3D QAH phase for the lattice model (blue solid line), the continuum model (red dashed line), and the analytic results (black dotted line). For the longitudinal (transverse) conductivities, the analytic results are obtained for $\gamma=0$ ($\gamma={m_0a^2\over 2}$). 
Solid (dashed) lines in the inset to (a) represent the energy dispersion for $J=1$ along the $k_z$ direction with $k_x=0$ ($k_x={\pi\over a}$) and $k_y=0$. The left inset to (b) represents the energy dispersion for $J=2$ along the $k_x$ direction with $k_z={\pi\over a}$ and $k_y=0$, and the right inset to (b) shows an enlarged view in $\sigma_{xx}(\omega)$ near the interband transition. Arrows in the insets indicate interband transitions corresponding to the kink structures appearing in $\sigma_{xx}$ and $\sigma_{xy}$. 
%Note that for $J=2$, a Mexican hat structure appears in the energy dispersion, exhibiting a shifted interband peak with respect to the $\gamma=0$ result.  
Here, $m_z/m_0=-0.8$ and $k_{\rm c}=\pi/a$ are used for the calculation.}
\label{fig:QAH_optical_conductivity}
\end{figure}
%%%%%%%%%%%%%%%%%%%%%%%%%%%%%%%%%%%%%%%%%%%%%%%%%

For the NI phase ($\alpha>0$) and 3D QAH phase ($\alpha<0$), we obtain the leading-order $\omega$ dependence of longitudinal optical conductivities analytically assuming $\gamma=0$ in the vicinity of $\hbar\omega=2|\alpha|$,
\begin{subequations}\label{eq:conductivity_zz_insulating}
\begin{eqnarray}
\sigma_{xx}(\omega)&\sim&\left(\hbar \omega -2 |\alpha|\right)^{\frac{1}{2}}\Theta(\hbar\omega-2|\alpha|),\\
\sigma_{zz}(\omega)&\sim&\left(\hbar \omega -2 |\alpha| \right)^{\frac{1}{J}+\frac{3}{2}}\Theta(\hbar\omega-2|\alpha|).
\end{eqnarray}
\end{subequations}
%The full analytic results including higher order corrections in $\omega$ can be found in the Supplemental Material \cite{Supplemental}.
Note that similarly to the WSM phase, $\sigma_{xx}(\omega)$ has the same $\omega$ dependence regardless of the chirality index $J$, while $\sigma_{zz}(\omega)$ has different power-law exponents depending on $J$. Here the analytic results are obtained assuming $\gamma=0$ for simplicity, which is valid when the effect of the band distortion associated with nonzero $\gamma$ is small ($\gamma k_0^2\ll \varepsilon_0$ or $m_0\ll t_x, t_y$). As $\gamma k_0^2/\varepsilon_0$ increases, the power-law exponent deviates from the analytic expression in Eq.~(\ref{eq:conductivity_zz_insulating}) obtained assuming $\gamma=0$, and the derivation is more significant for $J=2$ than $J=1$ because the kinetic term associated with $J$ is comparable to the quadratic $\gamma$ term at low frequencies \cite{Supplemental}.

The transverse optical conductivities in the NI and 3D QAH phases up to second order in $\omega$ are given by
\begin{equation}\label{eq:conductivity_transverse_insulating_phase}
\sigma_{xy}(\omega)=\xi\sigma_{xy}^{\rm QAH}+\frac{e^2}{\hbar} B_{xy}\omega^2,
\end{equation}
where $\sigma_{xy}^{\rm QAH}=\frac{J e^2}{\hbar}\frac{k_{\rm c}}{2 \pi^2}$ and $\xi=0$ ($\xi=1$) for the NI (3D QAH) phase. The static part ($\omega=0$) in Eq.~(\ref{eq:conductivity_transverse_insulating_phase}) can be obtained after properly subtracting the residual term, because the static Hall conductivity for the continuum model is not properly regularized carrying an arbitrary residual value. Thus only the difference in this quantity between different electronic states is experimentally measurable, giving a quantized value in the 3D QAH phase while zero in the NI phase. In this sense, we choose the momentum cutoff along the $k_z$ direction as $k_{\rm c}=\pi/a$ so that the properly subtracted static Hall conductivity in the 3D QAH phase has the same quantized value as in the lattice model.
A detailed discussion on the regularization process and the expression for $B_{xy}$ can be found in the Supplemental Material \cite{Supplemental}. Note that for the transverse optical conductivities, we present analytic results with non-zero $\gamma$.

Figure \ref{fig:QAH_optical_conductivity} shows calculated optical conductivities for the $J=1$ and $J=2$ lattice and continuum models in the 3D QAH phase. If $\gamma=0$, the energy gap with a size of $2|\alpha|$ for both NI and 3D QAH phases leads to zero conductivity for frequencies $\hbar\omega<2|\alpha|$ due to the optical gap. Because of the non-zero $\gamma$, a Mexican hat structure appears in the 3D QAH phase (but not in the NI phase) if $\alpha<\alpha_{\rm c}=-{\varepsilon_0^2 \over 2\gamma k_0^2}$ for $J=1$, and if $\alpha<0$ for $J=2$ exhibiting a shifted interband peak with respect to the $\gamma=0$ result \cite{Supplemental}. For the $J=1$ lattice model in the 3D QAH phase, an additional kink structure appears at $\hbar\omega=2|m_z-t_z+2m_0|$ due to the interband transitions at local minima $(k_x,k_y,k_z)=(\pm {\pi\over a},0,0)$, $(0,\pm {\pi\over a},0)$, as shown in Fig.~\ref{fig:QAH_optical_conductivity}(a). 
%Note that the analytic results for the longitudinal conductivities presented here are obtained assuming $\gamma=0$ for simplicity, which is valid when the effect of the band distortion associated with nonzero $\gamma$ is small ($m_0\ll t_x, t_y$). 

%%%%%%%%%%%%%%%%%%%%%%%%%%%%%%%%%%%%%%%%%%%%%%%%%%
\begin{figure}[!htb]
\includegraphics[width=1\linewidth]{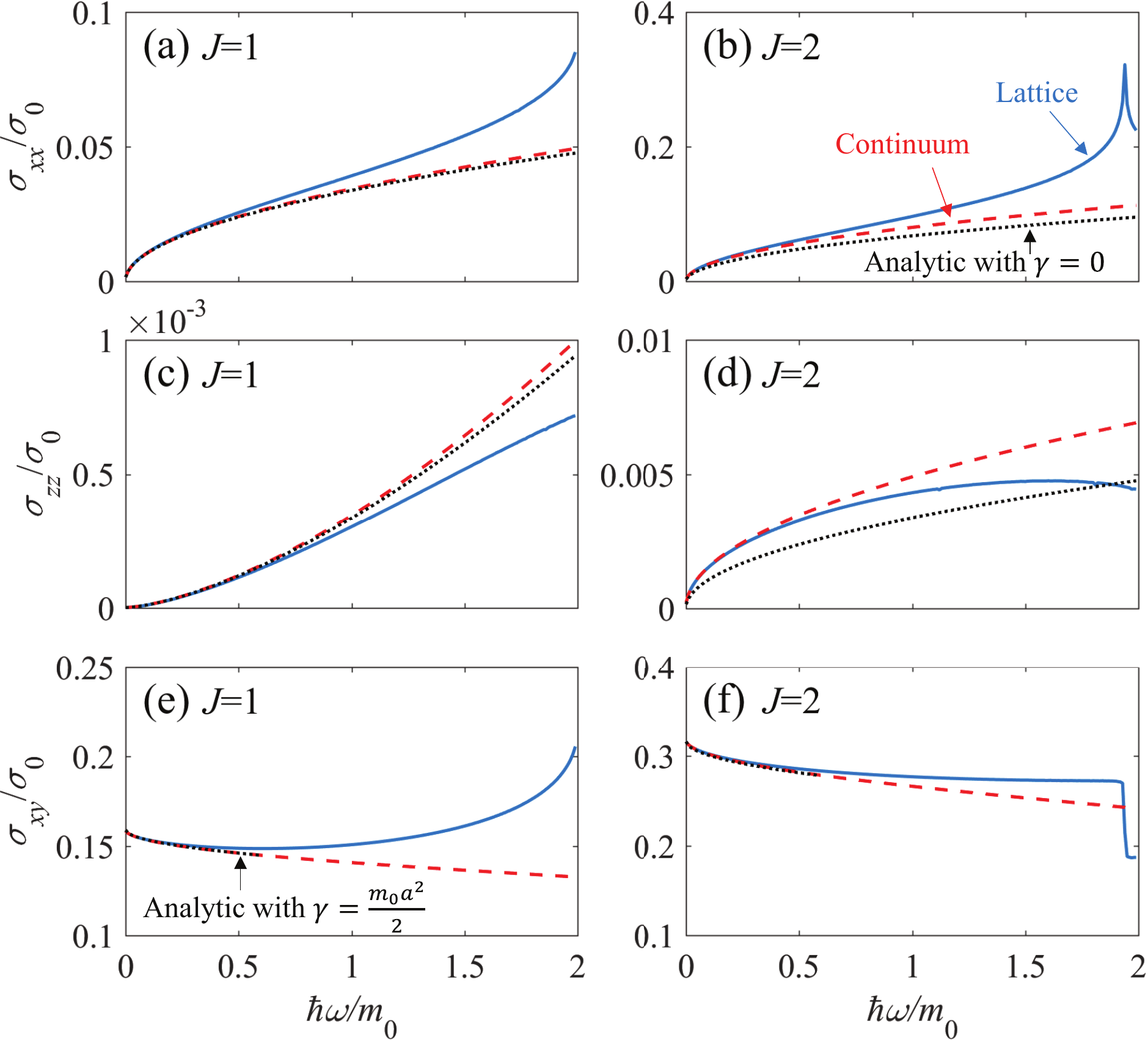}
\caption{
Real part of (a)-(d) longitudinal and (e), (f) transverse optical conductivities at the transition between the 3D QAH and WSM phases for the lattice model (blue solid line), the continuum model (red dashed line), and the analytic results (black dotted line).  For the longitudinal (transverse) conductivities, the analytic results are obtained for $\gamma=0$ ($\gamma={m_0a^2\over 2}$).  Here, $m_z/m_0=-0.5$ and $k_{\rm c}=\pi/a$ are used for calculation.
}
\label{fig:QAH_WSM_optical_conductivity}
\end{figure}
%%%%%%%%%%%%%%%%%%%%%%%%%%%%%%%%%%%%%%%%%%%%%%%%%%

% At the transition point between the WSM and NI phases or between the WSM and 3D QAH phases, the longitudinal (transverse) optical conductivities obtained assuming a zero (non-zero) $\gamma$ are given by
% \begin{subequations}\label{eq:conductivity_transition}
% \begin{eqnarray}
% \sigma_{xx}(\omega)&=&\frac{e^2}{\hbar} A_{xx} (\hbar\omega)^{1\over 2}, \\
% \sigma_{zz}(\omega)&=&\frac{e^2}{\hbar} A_{zz} (\hbar\omega)^{{2\over J}-{1\over 2}}, \\
% \sigma_{xy}(\omega)&=&\xi\sigma_{xy}^{\rm QAH}+\frac{e^2}{\hbar}\left[A_{xy}^{\rm NI}+B_{xy}\omega^{\nu}\right],
% \end{eqnarray}
% \end{subequations}
% where $\sigma_{xy}^{\rm QAH}=\frac{e^2}{\hbar} \frac{J k_{\rm c}}{2\pi^2}$ and $A_{xy}^{\rm NI}$ is a residual conductivity in Eq.~(\ref{eq:conductivity_transverse_insulating_phase_function}) for the NI phase that should be subtracted to obtain the absolute conductivity, which is consistent with the lattice model result. The expressions for $A_{xx}$ and $A_{zz}$ can be found in the Supplemental Material \cite{Supplemental}. Here, the exponent $\nu\approx 0.5$ is found numerically for $J=1, 2$ with a frequency independent coefficient $B_{xy}$. Note that the longitudinal conductivities for both transition points are identical (within a $\gamma=0$ approximation), whereas the Hall conductivities have different static values, with the difference given by $\sigma_{xy}^{\rm QAH}$.

At the transition point between the WSM and NI phases or between the WSM and 3D QAH phases, the longitudinal (transverse) optical conductivities obtained assuming a zero (non-zero) $\gamma$ are given by
\begin{subequations}\label{eq:conductivity_transition}
\begin{eqnarray}
\sigma_{xx}(\omega)&=&\frac{e^2}{\hbar} A_{xx} (\hbar\omega)^{1\over 2}, \\
\sigma_{zz}(\omega)&=&\frac{e^2}{\hbar} A_{zz} (\hbar\omega)^{{2\over J}-{1\over 2}}, \\
\sigma_{xy}(\omega)&=&\xi\sigma_{xy}^{\rm QAH}+\frac{e^2}{\hbar} C_{xy}\omega^{\nu}.
\end{eqnarray}
\end{subequations}
Note that similarly as in the NI and 3D QAH phases, the static part of $\sigma_{xy}(\omega)$ should be properly subtracted by the residual term. The expressions for $A_{xx}$ and $A_{zz}$ can be found in the Supplemental Material \cite{Supplemental}. Here, the exponent $\nu\approx 0.5$ is found numerically for $J=1, 2$ with a frequency independent coefficient $C_{xy}$. Note that the longitudinal conductivities for both transition points are identical (within a $\gamma=0$ approximation), whereas the Hall conductivities have different static values, with the difference given by $\sigma_{xy}^{\rm QAH}$.

%%%%%%%%%%%%%%%%%%%%%%%%%%%%%%%%%%%%%%%%%%%%%%%%%%%%%%%%%%%%%%%%%%%%%%%%%%%%%%%%%%%%%%%%%%%%%%%%%%%%
{\em Discussion} ---
Recently, Huang \textit{et al.} \cite{Huang2016} demonstrated that strontium silicide (SrSi$_2$) hosts double Weyl nodes with a chirality $J=2$.
The effective Hamiltonian, which describes one of the Weyl nodes with a chirality $J=2$ in SrSi$_2$, resembles that of bilayer graphene with the interlayer hopping replaced by the spin-orbit coupling $\Delta$ connecting the two $J=1$ Weyl Hamiltonians.
If we assume $\mu=0$, at low frequencies the optical conductivity for the double Weyl nodes in SrSi$_2$ behaves similar to that of the $J=2$ Weyl semimetals, showing $\sigma_{xx}\sim \omega$ and $\sigma_{zz}\sim \omega^0$ dependence, whereas at high frequencies, the optical conductivity shows two copies of the $J=1$ Weyl semimetals exhibiting a linear $\omega$ dependence in $\sigma_{xx}$ and $\sigma_{zz}$. At intermediate frequencies, kink structures appear at frequencies comparable to the energy scales of interband transitions determined by $\Delta$. Note that the double Weyl nodes in SrSi$_2$ are not actually located at $\mu=0$, thus the longitudinal conductivity in real SrSi$_2$ will give additional features of the Pauli blocking and the Drude peak. 
In addition, in a real sample, multiple Weyl nodes coexist, thus the optical conductivity can be obtained by the sum of the contribution from each node. Tilt and impurities will also affect the optical conductivity. However, the characteristic frequency dependence described here will not be altered above the frequency corresponding to the energy scale of the tilt or impurity potential \cite{Supplemental}.

In summary, we studied the optical properties of m-WSMs in semimetallic and nearby insulating phases, focusing on the frequency dependence of optical conductivity. We demonstrated that the optical conductivities $\sigma_{xx}(\omega)$, $\sigma_{zz}(\omega)$ and $\sigma_{xy}(\omega)$ show a characteristic frequency dependence that strongly varies according to the winding number and phase of the system, and thus can be used as a spectroscopic signature of m-WSMs.

\acknowledgments
This research was supported by the Basic Science Research Program through the National Research Foundation of Korea (NRF) funded by the Ministry of Education under Grant No. 2015R1D1A1A01058071. EJM's work on this project was supported by the U.S. Department of Energy, Office of Basic Energy Sciences under Award No. DE-FG02-ER45118. HM acknowledges travel support provided by the University Research Foundation at the University of Pennsylvania while this work was carried out. 

%%%%%%%%%%%%%%%%%%%%%%%%%%%%%%%%%%%%%%%%%%%%%%%%%%%%%%%%%%%%%%%%%%%%%%%%%%%%%%%%%%%%%%%%%%%%%%%%%%%%

%\vspace{20pt}

%\end{document}
%%%%%%%%%%%%%%%%%%%%%%%%%%%%%%%%%%%%%%%%%%%%%%%%%%

%%%%%%%%%%%%%%%%%%%%%%%%%%%%%%%%%%%%%%%%%%%%%%%%%%
\newpage
\pagenumbering{gobble}

\begin{figure}[htp]
\includegraphics[page=1,trim = 19mm 17mm 17mm 17mm]{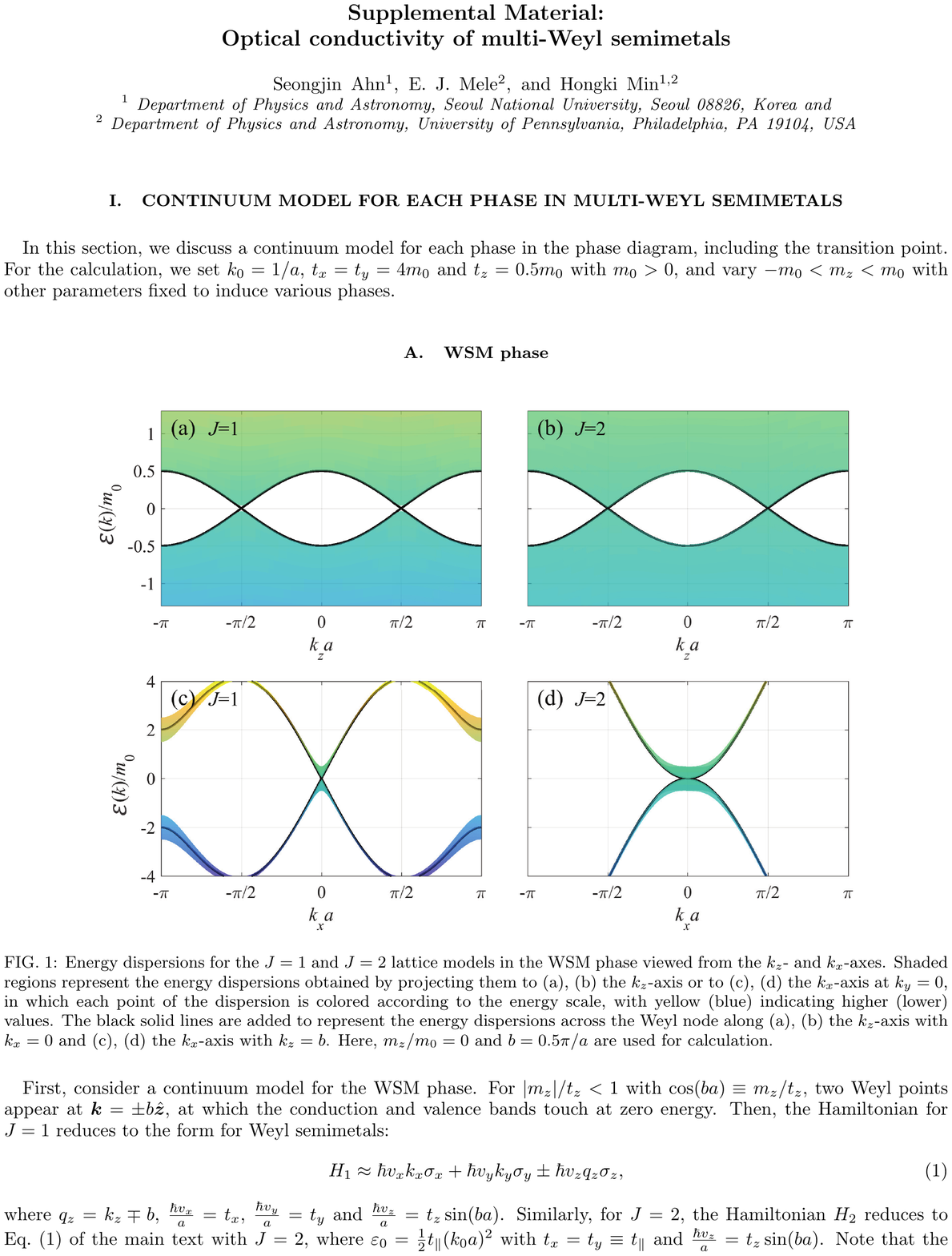}

\end{figure}

\newpage

\begin{figure}[htp]
  \includegraphics[page=2,trim = 19mm 17mm 17mm 17mm]{supplemental.pdf}

\end{figure}

\newpage

\begin{figure}[htp]
  \includegraphics[page=3,trim = 19mm 17mm 17mm 17mm ]{supplemental.pdf}

\end{figure}

\newpage

\begin{figure}[htp]
  \includegraphics[page=4,trim = 19mm 17mm 17mm 17mm ]{supplemental.pdf}

\end{figure}

\newpage

\begin{figure}[htp]
  \includegraphics[page=5,trim = 19mm 17mm 17mm 17mm ]{supplemental.pdf}

\end{figure}

\newpage

\begin{figure}[htp]
  \includegraphics[page=6,trim = 19mm 17mm 17mm 17mm ]{supplemental.pdf}

\end{figure}

\newpage

\begin{figure}[htp]
  \includegraphics[page=7,trim = 19mm 17mm 17mm 17mm ]{supplemental.pdf}

\end{figure}

\newpage

\begin{figure}[htp]
  \includegraphics[page=8,trim = 19mm 17mm 17mm 17mm ]{supplemental.pdf}

\end{figure}

\newpage

\begin{figure}[htp]
  \includegraphics[page=9,trim = 19mm 17mm 17mm 17mm ]{supplemental.pdf}

\end{figure}

\newpage

\begin{figure}[htp]
  \includegraphics[page=10,trim = 19mm 17mm 17mm 17mm ]{supplemental.pdf}

\end{figure}

\newpage

\begin{figure}[htp]
  \includegraphics[page=11,trim = 19mm 17mm 17mm 17mm ]{supplemental.pdf}

\end{figure}

\newpage

\begin{figure}[htp]
  \includegraphics[page=12,trim = 19mm 17mm 17mm 17mm ]{supplemental.pdf}

\end{figure}

\newpage

\begin{figure}[htp]
  \includegraphics[page=13,trim = 19mm 17mm 17mm 17mm ]{supplemental.pdf}

\end{figure}

\newpage

\begin{figure}[htp]
  \includegraphics[page=14, trim = 19mm 17mm 17mm 17mm ]{supplemental.pdf}

\end{figure}

\newpage

\begin{figure}[htp]
  \includegraphics[page=15,trim = 19mm 17mm 17mm 17mm ]{supplemental.pdf}

\end{figure}

\end{document}